\documentclass[aps,prb,superscriptaddress,twocolumn]{revtex4-1}
\usepackage{bm}
\usepackage[fleqn]{amsmath}
\usepackage{hyperref}
\usepackage{graphicx}

\begin{document}

\title{Collective excitations in two-component one-dimensional massless Dirac plasma}

\author{V.V. Enaldiev}
\email{vova.enaldiev@gmail.com}
\affiliation{Kotel'nikov Institute of Radio-engineering and Electronics of the Russian Academy of Sciences, 11-7 Mokhovaya St, Moscow, 125009 Russia}

\date{\today}

\begin{abstract}
We study spectra of long wavelength plasma oscillations in a system of two energy splitted one-dimensional (1D) massless Dirac fermion subbands coupled by spin-orbit interaction. Such a system may be formed by edge subbands in semiconducting transition metal dichalcogenide monolayers. Intrasubband transitions of massless Dirac fermions give rise to optical and acoustic gapless branches of intrasubband 1D plasmons. We reveal that the optical branch is of quantum character with group velocity being inverse proportional to square root of the Planck constant, whereas the acoustic branch is classical one with group velocity proportional to geometric mean of the edge subband velocities. Spin-orbit interaction, allowing intersubband transitions in the system, results in emergence of two branches of intersubband 1D plasmons: upper and lower ones. The upper and lower branches are gapped at small wave vectors and evolve  with positive and negative group velocities, respectively, from energy splitting of the edge subbands at Fermi-level. The both intersubband branches adjoin intersubband single particle excitation continuum from above, while in case of the edge subbands with unequal velocities the lower one experiences Landau damping at small wave vectors. In addition, the lower branch, attaining zero frequency at a non-zero wave vector, alters its group velocity from negative to positive one.
\end{abstract}

\maketitle

{\em Introduction.} Nowadays studies of systems with massless or massive Dirac fermions, such as graphene, topological insulators, monolayers of semiconducting transition metal dichalcogenides (sTMDs), attracts a lot of interest for both fundamental physics and application prospects in nanoelectronics and nanophotonics \cite{Franz2013, Novoselov2016}. An interesting feature of the Dirac fermion systems is that they have surface (in 3D) or edge (in 2D) states with a linear dispersion law \cite{VolkovJETP}. 
In the 2D case, the edge states (ESs) form 1D conducting channels of massless Dirac Fermions (MDFs) that freely propagate along the edge of the system. Despite in graphene manifestation of the ESs is masked by virtue of zero band gap, recent experiments on conductivity of gapped graphene are interpreted with help of the ESs \cite{Zhu2017}. ES conductance was established in 2D topological insulators realized in HgTe/CdTe/HgTe quantum wells of definite thickness \cite{Molenkamp2009}, but quite small band gap (about 20 meV) prevents them from using in devices at room temperatures. 

From the other hand, atomically thin sTMDs, such as MoS$_2$, WS$_2$, represent a 2D systems with massive Dirac fermions in the bulk \cite{bib:Xiao} and large band gap about 1.5-2 eV. Such a large band gap raises relevant question about ESs existing inside of it. Indeed, conducting ESs in the band gap was observed in scanning tunneling spectroscopy \cite{bib:Zhang} and
microwave impedance microscopy \cite{bib:Wua} of MoS$_2$ monolayers. Recently, transport features of edge states lying in the gap were observed in monolayers of 1T'-WTe$_2$ at temperatures up to 100K \cite{Fei2017,Wu2018}. 

In a minimal continuum model of sTMD monolayers, described by massive Weyl equation \cite{bib:Xiao}, theory predicts two energy-splitted subbands of ESs with linear spectra characterized by phenomenological parameters in a boundary condition \cite{bib:Korman, Enaldiev}. These ES subbands are spin-polarized in absence of an additional spin-orbit interaction (SOI) that entangles states with opposite spins. From collective excitations point of view, the ES subbands form a two-component 1D massless Dirac plasma. 

In this paper, we find spectra of plasma oscillations in such a system taking also into account additional SOI mechanisms. In the sTMD monolayers there are two types of the additional SOI that can couple the edge subbands characterized by opposite spin projections. The first type is related to the entanglement in the bulk of the monolayers, which can be induced by perpendicular electric field \cite{Basko}, Bychkov-Rashba SOI with the substrate \cite{kormanyos2014,Roldan} or in-plane magnetic field. The second type is induced by breaking down of periodic crystal potential at the edge \cite{Enaldiev}. 

We show that in the system under consideration there exist four branches of the 1D plasmons: two intrasubband branches (optical and acoustic), and two intersubband ones (upper and lower). We find that spectra of optical and acoustic intrasubband 1D plasmons being determined by one dispersion equation, unexpectedly, have essentially different dependence on the Planck constant. In the long wavelength limit frequencies of the optical branch are directly proportional to $1/\sqrt{\hbar}$, whereas frequencies of acoustic branch does not depend on the Planck constant. We also demonstrate that at small wave vectors upper and lower intersubband branches of 1D plasmons possess gapped spectra with positive and negative group velocities, respectively. The gap equals energy splitting of the edge subbands at the Fermi-level. In addition, we reveal that the lower branch reaching zero frequency with negative group velocity alters its sign to positive with increase of wave vector values.

{\em Model.--} We consider semi-infinite plane $y>0$ occupied by a sTMD monolayer. In the bulk, dynamics of carriers in the $K$ and $K'$ valleys of the sTMD monolayer is described by a massive 2D Weyl Hamiltonian \cite{bib:Xiao}:
\begin{equation}\label{TMD_Hamiltonian}
H_{\tau} = v\left(\tau\sigma_xp_x + \sigma_yp_y\right) +\sigma_z\left[m - \tau \hat{s}_z(\Delta_v-\Delta_c)\right]+ \sigma_0 \tau \hat{s}_z(\Delta_v+\Delta_c), 
\end{equation}
where $\tau$ is a valley index equal to $+1$ ($-1$) in the $K$ ($K'$) valley, $\sigma_{x,y,z}$ and $\sigma_0$ are Pauli and identity matrices, respectively, acting in band subspace, $\hat{s}_z$ is $z$ component of the spin-1/2 operator, $v$ is matrix element of momentum between the band extremum wave functions, $2\Delta_{c,v}$ is atomic spin-orbit splitting in conduction and valence bands, respectively, $2m$ is band gap without the splitting. The translational invariant edge along the $x$ axis is introduced through a boundary condition (BC) \cite{bib:Korman,Enaldiev}. Below we will consider zigzag orientation of the edge or close to it, which is more frequently realized in experiments \cite{Zande}. At the edge orientation, projections of the $K$ and $K'$ valleys onto the edge direction are well distant in reciprocal space.  Therefore, we will neglect intervalley coupling in BC. In addition, we will employ two types of BC depending on the SOI mechanisms. For SOI mechanisms of the first type, described by a perturbation term added to the Hamiltonian $H_{\tau}$ \cite{Basko, kormanyos2014, Roldan}, we choose spin-decoupled BC \cite{Enaldiev}: $\left[\psi_{c,\tau}^{\uparrow (\downarrow)} + a_{1,2;\tau} \psi_{v,\tau}^{\uparrow (\downarrow)}\right]_{x=0}=0$, where $a_{1,\tau}$ ($a_{2,\tau}$) is a real phenomenological parameter, characterizing edge properties for spin up (spin down) states, $\psi_{c,\tau}^{\uparrow (\downarrow)}$, $\psi_{v,\tau}^{\uparrow (\downarrow)}$ are envelope wave functions, describing spin up (spin down) states in conduction and valence bands, respectively. Time reversal symmetry implies the following relation for the parameters: $a_{1,\tau}=a_{2,-\tau}$. The SOI mechanism of the second type is described by a BC that entangles wave functions of opposite spin projections \cite{Enaldiev}:
$\left[\Psi^{\uparrow}_{\tau} - M_{\tau}\Psi^{\downarrow}_{\tau} \right]_{\rm edge} = 0,$ where $\Psi^{\uparrow(\downarrow)}_{\tau} = (\psi_{c,\tau}^{\uparrow (\downarrow)},\psi_{v,\tau}^{\uparrow (\downarrow)})$ is two-component envelope functions, $M_{\tau}$ is a second order matrix describing the edge SOI (explicit form of $M_{\tau}$ can be found in \cite{SM}). 

\begin{figure}
\includegraphics[scale=1.0]{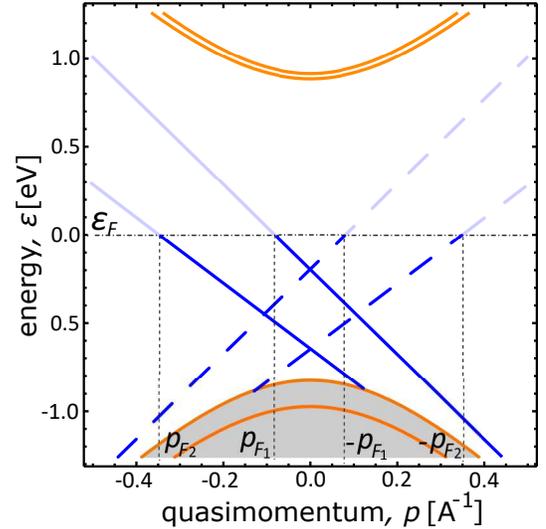}
\caption{ ES spectra (straight blue lines) (\ref{ES_spectra}) in $K$ (solid) and $K'$ (dashed) valleys. Solid gold lines mark edges of conduction and valence bands. The edge subbands cross Fermi-level (dashed-dotted line) at $p_{F_{1,2}}$ ($-p_{F_{1,2}}$) in the $K$ ($K'$) valley. Filled grey region indicates occupied states in valence band. Figure is plotted at the following bulk parameters: $2m=1.8$ eV, $v=2.5$ eV$\cdot$A, $2\Delta_c=3$ meV, $2\Delta_v=0.148$ eV. Parameters of ES spectra are the following: $v_1=0.8 v$, $v_2=0.62 v$, $\Delta_1=-0.23$ eV, $\Delta_2=-0.61$ eV. \label{Fig1} }
\end{figure}

Both types of the SOIs result in two subbands of ESs with linear spectra \cite{SM}:
\begin{equation}\label{ES_spectra}
\varepsilon^{(\tau)}_{n,p} = -\tau v^{(\tau)}_{n}p + \Delta^{(\tau)}_{n}, 
\end{equation}
where $n=1,2$ is the subband number, $p$ is quasimomentum along the edge, $v_{n}^{(\tau)}$ and $\Delta_{n}^{\tau}$ are velocity and value of electron-hole asymmetry in each ES subband, which are determined by bulk and boundary parameters. The ESs exist until their decay lengths are positive, turning into bulk states at those quasimomenta when their spectra (\ref{ES_spectra}) overlap with bulk continuum. Following experimental data \cite{bib:Wua,bib:Zhang,Wu2018} and tight-binding calculations \cite{bib:Guinea} we assume that ES subbands are in the band gap as shown on Fig.\ref{Fig1}. The upper edge subband $n=1$ ($n=2$)  always has greater velocity than the lower one $n=2$ ($n=1$) in the valley $K$ ($K'$), excluding their intersection. Time reversal symmetry leads to relations: $v_1^{(\tau)}=v_2^{(-\tau)}$, $\Delta_{1}^{(\tau)}=\Delta_{2}^{(-\tau)}$.  

{\em 1D Plasmon spectra.--} Introducing the single particle model, we turn to the question of collective excitation spectra in the two partly filled ES subbands (see Fig.\ref{Fig1}). In self-consistent random phase approximation, matrix elements of electric potential Fourier harmonic $\varphi_q(y)$, induced by density fluctuations of 1D MDF in the edge subbands, satisfy the following equation \cite{SM}:
\begin{eqnarray}\label{RPA_gen}
\langle \mu'|\varphi_{q}(y)e^{iqx}|\mu\rangle\delta_{\tau_{\mu'}\tau_{\mu}} = \qquad\qquad\qquad\qquad\qquad\qquad\qquad\nonumber\\
 \frac{2e^2}{\epsilon^*L_x}\sum_{\lambda\lambda'}\frac{f(\varepsilon_{\lambda'}) - f(\varepsilon_{\lambda}) }{ \varepsilon_{\lambda'} - \varepsilon_{\lambda} + \hbar\omega +i0 }\langle \lambda|\varphi_{q}(y)e^{iqx}|\lambda'\rangle\delta_{\tau_{\lambda'}\tau_{\lambda}}\times \nonumber\\ 
\langle\lambda'|\langle\mu'|e^{iq(x-x')}K_0\left(|q(y-y')|\right)|\mu\rangle|\lambda\rangle \qquad
\end{eqnarray}
here $\lambda,\mu$ denote sets of the ES quantum numbers ($n_{\lambda,\mu}=\{1,2\}$ is the subband number, $p_{\lambda,\mu}$ is quasimomentum along the edge, $\tau_{\lambda,\mu}$ is the valley index), $f(\varepsilon)$ is the Fermi-Dirac distribution function, $K_0(x)$ is the Macdonald function describing the Coulomb interaction in 1D system, $e$ is the absolute value of electron charge, $\epsilon^*$ is dielectric constant of environment, $L_x$ is size of the system in $x$ direction. The Kronecker delta in Eq.(\ref{RPA_gen}) means that we consider only intravalley transitions.

First, we take into account only diagonal  matrix elements on the subband number in Eq.(\ref{RPA_gen}) and obtain the dispersion equation for intrasubband 1D plasmons in the long wavelength limit $|q|\kappa_{\rm intra}^{-1}\ll 1$ (see \cite{SM}):
\begin{eqnarray}\label{intra_dispeq}
1-\frac{2e^2}{\epsilon^*}\ln\left(\frac{4\kappa_{\rm intra}}{|q|\zeta}\right)\sum_{\tau=\pm 1}\left[\Pi^{(\tau)}_{11}(\omega, q) + \Pi^{(\tau)}_{22}(\omega, q)\right]=0, \qquad\\
\Pi^{(\tau)}_{n'n}(\omega, q)=\frac{1}{L_x}\sum_{p}\frac{f\left(\varepsilon_{n',p}^{(\tau)}\right) - f\left(\varepsilon_{n,p+q}^{(\tau)}\right) }{ \varepsilon_{n',p}^{(\tau)} - \varepsilon_{n,p+q}^{(\tau)} + \hbar\omega + i0 }, \, n,n'=\{1,2\} \nonumber
\end{eqnarray}  
where $\zeta=e^{\gamma}$, $\gamma=0.577\dots$ is the Euler constant, \mbox{$\kappa_{\rm intra}$} is a characteristic intrasubband inverse decay length of the ES wave functions at Fermi-energy (expression for $\kappa_{\rm intra}$ through the model parameters can be found in \cite{SM}).

\begin{figure}
	\includegraphics{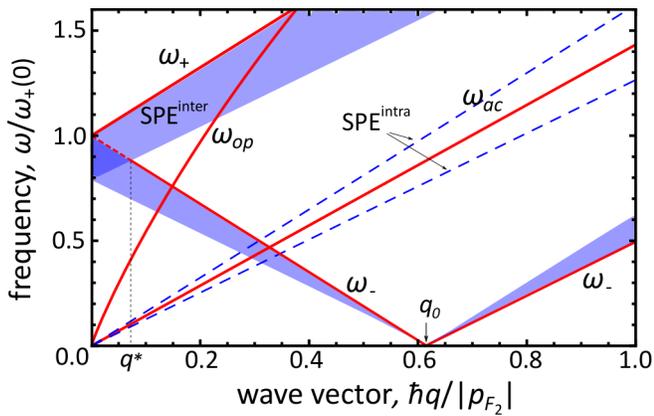}
	\caption{Red solid lines show wave vector dispersion of plasmons in two-component 1D massless Dirac plasma in case of unequal velocities of the edge subbands. $\omega_{op}$ and $\omega_{ac}$ identify spectra of intrasubband optical (\ref{intra_optic}) and acoustic (\ref{intra_acoustic}) branches, respectively. Blue dashed lines indicate intrasubband SPE continuum: \mbox{$\omega=v_1q$} and \mbox{$\omega=v_2q$}. Intersubband plasmons $\omega_{+,-}$ (\ref{inter_plsmn_v1neqv2}) exponentially adjoin the intersubband SPE continuum (blue shaded area). The lower intersubband branch $\omega_-$ also lies in the intersubband SPE continuum at \mbox{$|q|<q^*=\delta v\delta p_F/\hbar(v_1+v_2)$}. At $q_0=(p_{F_1}-p_{F_2})/\hbar$ the lower branch alters its group velocity from $-v_1$ to $v_2$. Calculation was carried out at $v_1=0.8 v$, $v_2=0.62 v$, $\Delta_1=-0.48$ eV, $\Delta_2=-0.59$ eV, $\varepsilon_F=-0.2$ eV, the bulk parameters as on Fig.\ref{Fig1}. \label{Fig2}}
\end{figure}

Solving Eq.~(\ref{intra_dispeq}) in the temperature range \mbox{$T\ll m-\varepsilon_F$} we obtain spectra of optical and acoustic branches of intrasubband 1D plasmons at $q\to 0$:
\begin{equation}\label{intra_optic}
\omega_{op}(q) \approx q\left(v_1+v_2\right)\sqrt{\frac{2}{\pi}\alpha\ln\left(\frac{4\kappa_{\rm intra}}{|q|\zeta}\right)}
\end{equation}
\begin{equation}\label{intra_acoustic}
\omega_{ac}(q) \approx q\sqrt{v_{1}v_{2}},
\end{equation}
where \mbox{$v_{1,2}\equiv v_{1,2}^{(1)}$}, \mbox{$\alpha=e^2/\epsilon^*\hbar(v_1+v_2)$} is the effective fine structure constant. It is Coulomb interaction of 1D MDFs filling two edge subbands that couples the two intrasubband 1D plasmons, resulting in optical and acoustic branches (\ref{intra_optic}),(\ref{intra_acoustic}).

Frequencies of the optical branch only logarithmically depend on the edge subband filling via $\kappa_{\rm intra}$ taken at Fermi-level \cite{Volkov_Zag}. Furthermore, in leading order on $q$ dispersion of the optical branch are directly proportional to \mbox{$\sqrt{\alpha}\sim 1/\sqrt{\hbar}$} making it clearly quantum \cite{DasSarma}. The reason of both the unusual things is that ''effective masses'' of MDFs introduced as $m^*_{1,2}\equiv |p_{F_{1,2}}-p_{e_{1,2}}|/v_{1,2}$ are proportional to the concentration of 1D MDFs in the edge subbands $n^{(1,2)}_{\rm 1D}=|p_{F_{1,2}}-p_{e_{1,2}}|/\hbar\pi$. Therefore, in a standard formula for optical 1D plasmons of two-component plasma of massive fermions \cite{Wendler1994}:
\begin{eqnarray}\label{common_intra_plasmon}
\widetilde{\omega}_{op}(q)\approx q\sqrt{\frac{2e^2}{\epsilon^*}\left(\frac{n^{(1)}_{\rm 1D}}{m^*_{1}} + \frac{n^{(2)}_{\rm 1D}}{m^*_{2}}\right) \ln\left(\frac{4\kappa_{\rm intra}}{|q|\zeta}\right)}
\end{eqnarray}
one obtains unconventional shortening of 1D concentrations with the ''effective masses'', and as a consequence  proportionality to $\sqrt{\alpha}$ in Eq.(\ref{intra_optic}). In case of equal velocities of the edge subbands $v_1=v_2$, the optical branch frequencies are $\sqrt{2}$ times larger those of the 1D plasmons in one-component massless Dirac plasma \cite{DasSarma,Volkov_Zag}. 

The acoustic branch (\ref{intra_acoustic}) possesses linear dispersion at small wave vectors with group velocity proportional to geometric mean of the edge subband velocities. In the limit the branch does not depend on filling of the edge subbands and dielectric environment. Moreover, unlike the optical branch the acoustic one is non quantum as its dispersion (\ref{intra_acoustic}) does not contain the Planck constant. 

\begin{figure}
	\includegraphics{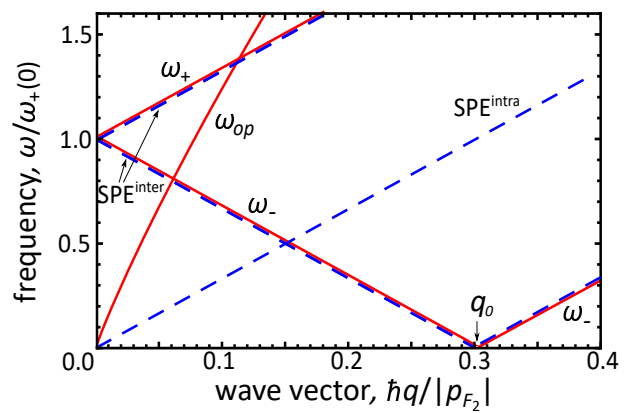}
	\caption{ Red solid lines shows wave vector dispersion of plasmons in two-component 1D massless Dirac plasma in case of equal velocities of the edge subbands (i.e. \mbox{$v_1=v_2$}). Only optical intrasubband branch $\omega_{op}$ (\ref{intra_optic}) exists in this case. Blue dashed lines indicate both intra- and intersubband SPE continua. Intersubband branches $\omega_{+,-}$ (\ref{inter_plsmn_v1=v2}) quadratically ((i.e. \mbox{$\propto q^2\ln(4\kappa_{\rm inter}/q\zeta )$})) deviate from the intersubband SPE continuum. At $q_0$ the lower branch changes the sign of its group velocity. Calculation was carried out at $v_1=v_2=0.8 v$, $\Delta_1=-0.48$ eV, $\Delta_2=-0.59$ eV, $\varepsilon_F=-0.2$ eV, the bulk parameters as on Fig.\ref{Fig1}. \label{Fig3}}
\end{figure}

Since continuum of intrasubband single particle excitations (SPEs) in the 1D system with linear spectra represents straight lines $\omega=v_{1}q$, $\omega=v_2q$ (blue dashed lines on Fig.\ref{Fig2},\ref{Fig3}), both intrasubband plasmons (\ref{intra_optic}),(\ref{intra_acoustic}) avoid Landau damping. However, at equal velocities of the edge subbands (i.e. \mbox{$v_1=v_2=v$}) acoustic branch (\ref{intra_acoustic}) disappears, as it coincides with intrasubband SPE continuum: $\omega=vq$ (see Fig.\ref{Fig3}). Here we note that  intrasubband 1D plasmons were also studied in density functional formalism\cite{Andersen}, where it was found three intrasubband branches as the model comprised three edge subbands.

Second, we consider intersubband plasmons. To obtain their spectra  one need to keep off diagonal matrix elements on subband index in  Eq.(\ref{RPA_gen}). In the long wavelength limit \mbox{$q\kappa_{\rm inter}^{-1}\ll 1$} we arrive to the following dispersion equation \cite{SM}:
\begin{eqnarray}\label{inter_disp}
1-\frac{2e^2}{\epsilon^*}\left|J_{12}^{(1)}(p_{F_2},q)\right|^2\ln\left ( \frac{4\kappa_{\rm inter}}{|q|\zeta} \right )\times \qquad\qquad\qquad\quad\nonumber \\
\sum_{\tau=\pm 1}\left[\Pi^{(\tau)}_{12}(\omega,q) + \Pi^{(\tau)}_{21}(\omega,q)\right] =0,\quad
\end{eqnarray}
where \mbox{$J_{12}^{(1)}(p_{F_2},q)\equiv \langle 1,p_{F_2}+q,1 |e^{iqx}|2,p_{F_2},1\rangle$} is matrix element of intersubband transitions that is non zero only due to presence of the additional SOI mechanisms discussed above (quasimomentum $p_{F_{2}}$ is determined by the equality \mbox{$\varepsilon_{2,p_{F_{2}}}^{(1)}=\varepsilon_F$}, see Fig.\ref{Fig1}), $\kappa_{\rm inter}$ is the intersubband inverse decay length at Fermi-level. In the long wavelength limit the matrix element is directly proportional to wave vector $q$ and  also is of weak dependence on quasimomentum $p$ (see \cite{SM}). Therefore, in derivation of Eq.(\ref{inter_disp}) we carried it out of the summation over momentum in Eq.(\ref{RPA_gen}) and took it at $p=p_{F_2}$. Below, we calculate intersubband plasmon dispersion at low temperature limit, when Fermi-Dirac distribution functions can be treated as step-like ones: \mbox{$f(\varepsilon_{1,2;p}^{(1)})=\theta[p-p_{F_{1,2}}]$},\mbox{ $f(\varepsilon_{1,2;p}^{(-1)})=\theta[-p-p_{F_{2,1}}]$}, $\theta[x]$ is the Heaviside function. In this case the intersubband polarization operator reads as follows:
\begin{eqnarray}\label{inter_polarization}
\Pi_{12}^{(\tau)}(\omega,q)=\frac{s}{2\hbar\delta v}\times\qquad\qquad\qquad\qquad\qquad\qquad\qquad\qquad\quad\nonumber \\
\left\{\frac{1}{\pi}\ln\left|\frac{\tau\hbar\omega + \Omega_2 }{\tau\hbar\omega + \Omega_1}\right| +
i\tau\theta\left[s\left(\tau\hbar\omega + \Omega_1\right)\right ]
\theta\left[s\left( -\tau\hbar\omega-\Omega_2\right)\right]\right\} \quad \\
 \Pi_{21}^{(\tau)}(\omega,q) = \left[\Pi_{12}^{(\tau)}(-\omega,-q)\right]^*\qquad\qquad\qquad\qquad\qquad\qquad\quad \nonumber 
\end{eqnarray}
where \mbox{$\delta v = v_1-v_2$}, \mbox{$s={\rm sgn}(p_{F_1}-p_{F_2}+\hbar q)$}, \mbox{$\Omega_{1,2}=\Delta_1-\Delta_2 + \hbar v_{1,2}q-\delta vp_{F_{2,1}}$}, (\mbox{$\Delta_{1,2}=\Delta^{(1)}_{1,2}$}). Substituting the formula (\ref{inter_polarization}) into Eq. (\ref{inter_disp}) we obtain two branches of intersubband 1D plasmons $\omega_{+}$ (upper) and $\omega_{-}$ (lower) shown on Figs.\ref{Fig2}, \ref{Fig3}). At $|q|<q_0\equiv (p_{F_1}-p_{F_2})/\hbar$ their spectra are determined by the following formula:
\begin{equation}\label{inter_plsmn_v1neqv2}
\omega_{\pm}(q) = \omega_{12} \pm v_1 q + 
\frac{\delta v q_0\left|1\pm \dfrac{\delta v q_0}{2v_1 q}\right|}{\exp\left[-\dfrac{\epsilon^*\pi\hbar\delta v}{e^2|J_{12}^{(1)}(p_{F_2},q)|^2\ln\left(\frac{4\kappa_{\rm inter}}{q\zeta}\right)}\right]-1}
\end{equation}  
where \mbox{$\hbar\omega_{12}=\varepsilon^{(1)}_{1,p_{F_2}}-\varepsilon^{(1)}_{2,p_{F_2}}$} is the gap in the plasmon spectra that is of the order of bulk band splitting (\mbox{$\sim 0.01-0.1$} eV).  The upper (lower) branch $\omega_{+}$ ($\omega_-$) (\ref{inter_plsmn_v1neqv2}) possesses positive $(\approx v_1)$ (negative $(\approx -v_1)$) group velocity and approaches boundary of intersubband SPE continuum from above. At $|q|<q^*=\delta vp_{F_2}/(v_1+v_2)\hbar$ the lower branch lies in the intersubband SPE continuum, leaving the latter at larger wave vectors ($|q|>q^*$). This differs from 1D systems of massive fermions, where both intersubband plasmons lie outside intersubband SPE in the long wavelength limit \cite{Wendler1994, Chaplik}. Frequency of the lower branch $\omega_{-}$ equals zero at $q = q_0$. At larger wave vectors ($q > q_0$) its spectrum reads as follows:
\begin{equation}\label{lower_branch_q>q0}
\omega_{-}(q)=v_2(q-q_0) - \frac{\delta v q_0\left|1\pm \dfrac{\delta v q_0}{2 v_1 q}\right|}{\exp\left[-\dfrac{\epsilon^*\pi\hbar\delta v}{e^2|J_{12}^{(1)}(p_{F_2},q)|^2\ln(4\kappa_{\rm inter}/q\zeta)}\right]-1}.
\end{equation}
Thus, at $q > q_0$ the lower branch has positive group velocity $(\approx v_2)$ and adjoins the bottom of the intersubband SPE continuum (see Figs.\ref{Fig2}, \ref{Fig3}).

Until the ES subbands have unequal velocities (i.e. $v_1 \neq v_2$) the intersubband plasmons (\ref{inter_plsmn_v1neqv2}) exponentially close adjoin the upper boundary of the intersubband SPE continuum as $J_{12}^{(1)}(p_{F_2},q)\propto q$. However at equal velocities ($v_1=v_2$) from Eq.(\ref{inter_plsmn_v1neqv2}) we obtain the following expression for 1D intersubband plasmons:
\begin{equation}\label{inter_plsmn_v1=v2}
\omega_{\pm}(q) = \omega_{12} \pm v q + \frac{e^2 q_{0}}{\epsilon^*\pi\hbar}|J_{12}^{(1)}(p_{F_2},q)|^2\ln\left(\frac{4\kappa_{\rm inter}}{q\zeta}\right). 
\end{equation}   
In this case both upper and lower intersubband plasmon branches (\ref{inter_plsmn_v1=v2}) only parabolically approach the intersubband SPE continuum represented by dashed blue lines $\omega=\omega_{12} \pm vq$ on Fig.\ref{Fig3}. It is of note that intersubband plasmons (\ref{inter_plsmn_v1neqv2}),(\ref{inter_plsmn_v1=v2}) have no polarization shift unlike intersubband 1D plasmons in a system of massive fermions \cite{Wendler1994, Chaplik}. 

In summary, we analytically derived long wavelength spectra of plasmons in two-component 1D massless Dirac plasma that is realized in system of two edge subbands of sTMD monolayers. The spectra consist of the two (optical and acoustic) intrasubband and two (upper and lower) intersubband branches. We found that the optical and acoustic branches are both gapless but have different quantum nature. The optical plasmon is quantum because of explicit dependence of its spectrum on the Planck constant, which follows from relation between 1D MDF concentration and their ''effective masses'' at Fermi-level. While the acoustic branch is rather classical as its group velocity is only determined by geometric mean of the edge subband velocities. We revealed that interubband branches arose due to intersubband transitions provided by the additional SOIs that can be either bulk or edge in nature. The upper and lower intersubband branches are gapped and possess positive and negative group velocities, respectively, at small wave vectors. We demonstrated that the lower branch, reaching zero frequency, changes its group velocity from negative  to positive ones. At the end we note that the developed theory of 1D plasmons can be particularly useful for understanding of response of sTMD monolayers at sub-gap frequencies.

\acknowledgements
This work was supported by the Russian Science Foundation (project no. 18-72-00022).
\bibliography{Bibl}

\begin{thebibliography}{23}%
\makeatletter
\providecommand \@ifxundefined [1]{%
 \@ifx{#1\undefined}
}%
\providecommand \@ifnum [1]{%
 \ifnum #1\expandafter \@firstoftwo
 \else \expandafter \@secondoftwo
 \fi
}%
\providecommand \@ifx [1]{%
 \ifx #1\expandafter \@firstoftwo
 \else \expandafter \@secondoftwo
 \fi
}%
\providecommand \natexlab [1]{#1}%
\providecommand \enquote  [1]{``#1''}%
\providecommand \bibnamefont  [1]{#1}%
\providecommand \bibfnamefont [1]{#1}%
\providecommand \citenamefont [1]{#1}%
\providecommand \href@noop [0]{\@secondoftwo}%
\providecommand \href [0]{\begingroup \@sanitize@url \@href}%
\providecommand \@href[1]{\@@startlink{#1}\@@href}%
\providecommand \@@href[1]{\endgroup#1\@@endlink}%
\providecommand \@sanitize@url [0]{\catcode `\\12\catcode `\$12\catcode
  `\&12\catcode `\#12\catcode `\^12\catcode `\_12\catcode `\%12\relax}%
\providecommand \@@startlink[1]{}%
\providecommand \@@endlink[0]{}%
\providecommand \url  [0]{\begingroup\@sanitize@url \@url }%
\providecommand \@url [1]{\endgroup\@href {#1}{\urlprefix }}%
\providecommand \urlprefix  [0]{URL }%
\providecommand \Eprint [0]{\href }%
\providecommand \doibase [0]{http://dx.doi.org/}%
\providecommand \selectlanguage [0]{\@gobble}%
\providecommand \bibinfo  [0]{\@secondoftwo}%
\providecommand \bibfield  [0]{\@secondoftwo}%
\providecommand \translation [1]{[#1]}%
\providecommand \BibitemOpen [0]{}%
\providecommand \bibitemStop [0]{}%
\providecommand \bibitemNoStop [0]{.\EOS\space}%
\providecommand \EOS [0]{\spacefactor3000\relax}%
\providecommand \BibitemShut  [1]{\csname bibitem#1\endcsname}%
\let\auto@bib@innerbib\@empty
\bibitem [{\citenamefont {Franz}\ and\ \citenamefont
  {Molenkamp}(2013)}]{Franz2013}%
  \BibitemOpen
  \bibfield  {author} {\bibinfo {author} {\bibfnamefont {M.}~\bibnamefont
  {Franz}}\ and\ \bibinfo {author} {\bibfnamefont {L.}~\bibnamefont
  {Molenkamp}},\ }\href@noop {} {\emph {\bibinfo {title} {Topological
  Insulators}}},\ Vol.~\bibinfo {volume} {6}\ (\bibinfo  {publisher}
  {Elsevier},\ \bibinfo {year} {2013})\BibitemShut {NoStop}%
\bibitem [{\citenamefont {Novoselov}\ \emph {et~al.}(2016)\citenamefont
  {Novoselov}, \citenamefont {Mishchenko}, \citenamefont {Carvalho},\ and\
  \citenamefont {Neto}}]{Novoselov2016}%
  \BibitemOpen
  \bibfield  {author} {\bibinfo {author} {\bibfnamefont {K.}~\bibnamefont
  {Novoselov}}, \bibinfo {author} {\bibfnamefont {A.}~\bibnamefont
  {Mishchenko}}, \bibinfo {author} {\bibfnamefont {A.}~\bibnamefont
  {Carvalho}}, \ and\ \bibinfo {author} {\bibfnamefont {A.~C.}\ \bibnamefont
  {Neto}},\ }\href@noop {} {\bibfield  {journal} {\bibinfo  {journal}
  {Science}\ }\textbf {\bibinfo {volume} {353}},\ \bibinfo {pages} {aac9439}
  (\bibinfo {year} {2016})}\BibitemShut {NoStop}%
\bibitem [{\citenamefont {Volkov}\ and\ \citenamefont
  {Enaldiev}(2016)}]{VolkovJETP}%
  \BibitemOpen
  \bibfield  {author} {\bibinfo {author} {\bibfnamefont {V.}~\bibnamefont
  {Volkov}}\ and\ \bibinfo {author} {\bibfnamefont {V.}~\bibnamefont
  {Enaldiev}},\ }\href@noop {} {\bibfield  {journal} {\bibinfo  {journal}
  {Journal of Experimental and Theoretical Physics}\ }\textbf {\bibinfo
  {volume} {122}},\ \bibinfo {pages} {608} (\bibinfo {year}
  {2016})}\BibitemShut {NoStop}%
\bibitem [{\citenamefont {Zhu}\ \emph {et~al.}(2017)\citenamefont {Zhu},
  \citenamefont {Kretinin}, \citenamefont {Thompson}, \citenamefont {Bandurin},
  \citenamefont {Hu}, \citenamefont {Yu}, \citenamefont {Birkbeck},
  \citenamefont {Mishchenko}, \citenamefont {Vera-Marun}, \citenamefont
  {Watanabe} \emph {et~al.}}]{Zhu2017}%
  \BibitemOpen
  \bibfield  {author} {\bibinfo {author} {\bibfnamefont {M.}~\bibnamefont
  {Zhu}}, \bibinfo {author} {\bibfnamefont {A.}~\bibnamefont {Kretinin}},
  \bibinfo {author} {\bibfnamefont {M.~D.}\ \bibnamefont {Thompson}}, \bibinfo
  {author} {\bibfnamefont {D.}~\bibnamefont {Bandurin}}, \bibinfo {author}
  {\bibfnamefont {S.}~\bibnamefont {Hu}}, \bibinfo {author} {\bibfnamefont
  {G.}~\bibnamefont {Yu}}, \bibinfo {author} {\bibfnamefont {J.}~\bibnamefont
  {Birkbeck}}, \bibinfo {author} {\bibfnamefont {A.}~\bibnamefont
  {Mishchenko}}, \bibinfo {author} {\bibfnamefont {I.}~\bibnamefont
  {Vera-Marun}}, \bibinfo {author} {\bibfnamefont {K.}~\bibnamefont
  {Watanabe}},  \emph {et~al.},\ }\href@noop {} {\bibfield  {journal} {\bibinfo
   {journal} {Nature Communications}\ }\textbf {\bibinfo {volume} {8}},\
  \bibinfo {pages} {14552} (\bibinfo {year} {2017})}\BibitemShut {NoStop}%
\bibitem [{\citenamefont {Roth}\ \emph {et~al.}(2009)\citenamefont {Roth},
  \citenamefont {Br{\"u}ne}, \citenamefont {Buhmann}, \citenamefont
  {Molenkamp}, \citenamefont {Maciejko}, \citenamefont {Qi},\ and\
  \citenamefont {Zhang}}]{Molenkamp2009}%
  \BibitemOpen
  \bibfield  {author} {\bibinfo {author} {\bibfnamefont {A.}~\bibnamefont
  {Roth}}, \bibinfo {author} {\bibfnamefont {C.}~\bibnamefont {Br{\"u}ne}},
  \bibinfo {author} {\bibfnamefont {H.}~\bibnamefont {Buhmann}}, \bibinfo
  {author} {\bibfnamefont {L.~W.}\ \bibnamefont {Molenkamp}}, \bibinfo {author}
  {\bibfnamefont {J.}~\bibnamefont {Maciejko}}, \bibinfo {author}
  {\bibfnamefont {X.-L.}\ \bibnamefont {Qi}}, \ and\ \bibinfo {author}
  {\bibfnamefont {S.-C.}\ \bibnamefont {Zhang}},\ }\href@noop {} {\bibfield
  {journal} {\bibinfo  {journal} {Science}\ }\textbf {\bibinfo {volume}
  {325}},\ \bibinfo {pages} {294} (\bibinfo {year} {2009})}\BibitemShut
  {NoStop}%
\bibitem [{\citenamefont {Xiao}\ \emph {et~al.}(2012)\citenamefont {Xiao},
  \citenamefont {Liu}, \citenamefont {Feng}, \citenamefont {Xu},\ and\
  \citenamefont {Yao}}]{bib:Xiao}%
  \BibitemOpen
  \bibfield  {author} {\bibinfo {author} {\bibfnamefont {D.}~\bibnamefont
  {Xiao}}, \bibinfo {author} {\bibfnamefont {G.-B.}\ \bibnamefont {Liu}},
  \bibinfo {author} {\bibfnamefont {W.}~\bibnamefont {Feng}}, \bibinfo {author}
  {\bibfnamefont {X.}~\bibnamefont {Xu}}, \ and\ \bibinfo {author}
  {\bibfnamefont {W.}~\bibnamefont {Yao}},\ }\href@noop {} {\bibfield
  {journal} {\bibinfo  {journal} {Physical Review Letters}\ }\textbf {\bibinfo
  {volume} {108}},\ \bibinfo {pages} {196802} (\bibinfo {year}
  {2012})}\BibitemShut {NoStop}%
\bibitem [{\citenamefont {Zhang}\ \emph {et~al.}(2014)\citenamefont {Zhang},
  \citenamefont {Johnson}, \citenamefont {Hsu}, \citenamefont {Li},\ and\
  \citenamefont {Shih}}]{bib:Zhang}%
  \BibitemOpen
  \bibfield  {author} {\bibinfo {author} {\bibfnamefont {C.}~\bibnamefont
  {Zhang}}, \bibinfo {author} {\bibfnamefont {A.}~\bibnamefont {Johnson}},
  \bibinfo {author} {\bibfnamefont {C.-L.}\ \bibnamefont {Hsu}}, \bibinfo
  {author} {\bibfnamefont {L.-J.}\ \bibnamefont {Li}}, \ and\ \bibinfo {author}
  {\bibfnamefont {C.-K.}\ \bibnamefont {Shih}},\ }\href@noop {} {\bibfield
  {journal} {\bibinfo  {journal} {Nano letters}\ }\textbf {\bibinfo {volume}
  {14}},\ \bibinfo {pages} {2443} (\bibinfo {year} {2014})}\BibitemShut
  {NoStop}%
\bibitem [{\citenamefont {Wu}\ \emph {et~al.}(2016)\citenamefont {Wu},
  \citenamefont {Li}, \citenamefont {Luan}, \citenamefont {Wu}, \citenamefont
  {Li}, \citenamefont {Yogeesh}, \citenamefont {Ghosh}, \citenamefont {Chu},
  \citenamefont {Akinwande}, \citenamefont {Niu} \emph {et~al.}}]{bib:Wua}%
  \BibitemOpen
  \bibfield  {author} {\bibinfo {author} {\bibfnamefont {D.}~\bibnamefont
  {Wu}}, \bibinfo {author} {\bibfnamefont {X.}~\bibnamefont {Li}}, \bibinfo
  {author} {\bibfnamefont {L.}~\bibnamefont {Luan}}, \bibinfo {author}
  {\bibfnamefont {X.}~\bibnamefont {Wu}}, \bibinfo {author} {\bibfnamefont
  {W.}~\bibnamefont {Li}}, \bibinfo {author} {\bibfnamefont {M.~N.}\
  \bibnamefont {Yogeesh}}, \bibinfo {author} {\bibfnamefont {R.}~\bibnamefont
  {Ghosh}}, \bibinfo {author} {\bibfnamefont {Z.}~\bibnamefont {Chu}}, \bibinfo
  {author} {\bibfnamefont {D.}~\bibnamefont {Akinwande}}, \bibinfo {author}
  {\bibfnamefont {Q.}~\bibnamefont {Niu}},  \emph {et~al.},\ }\href@noop {}
  {\bibfield  {journal} {\bibinfo  {journal} {Proceedings of the National
  Academy of Sciences}\ }\textbf {\bibinfo {volume} {113}},\ \bibinfo {pages}
  {8583} (\bibinfo {year} {2016})}\BibitemShut {NoStop}%
\bibitem [{\citenamefont {Fei}\ \emph {et~al.}(2017)\citenamefont {Fei},
  \citenamefont {Palomaki}, \citenamefont {Wu}, \citenamefont {Zhao},
  \citenamefont {Cai}, \citenamefont {Sun}, \citenamefont {Nguyen},
  \citenamefont {Finney}, \citenamefont {Xu},\ and\ \citenamefont
  {Cobden}}]{Fei2017}%
  \BibitemOpen
  \bibfield  {author} {\bibinfo {author} {\bibfnamefont {Z.}~\bibnamefont
  {Fei}}, \bibinfo {author} {\bibfnamefont {T.}~\bibnamefont {Palomaki}},
  \bibinfo {author} {\bibfnamefont {S.}~\bibnamefont {Wu}}, \bibinfo {author}
  {\bibfnamefont {W.}~\bibnamefont {Zhao}}, \bibinfo {author} {\bibfnamefont
  {X.}~\bibnamefont {Cai}}, \bibinfo {author} {\bibfnamefont {B.}~\bibnamefont
  {Sun}}, \bibinfo {author} {\bibfnamefont {P.}~\bibnamefont {Nguyen}},
  \bibinfo {author} {\bibfnamefont {J.}~\bibnamefont {Finney}}, \bibinfo
  {author} {\bibfnamefont {X.}~\bibnamefont {Xu}}, \ and\ \bibinfo {author}
  {\bibfnamefont {D.~H.}\ \bibnamefont {Cobden}},\ }\href@noop {} {\bibfield
  {journal} {\bibinfo  {journal} {Nature Physics}\ }\textbf {\bibinfo {volume}
  {13}},\ \bibinfo {pages} {677} (\bibinfo {year} {2017})}\BibitemShut
  {NoStop}%
\bibitem [{\citenamefont {Wu}\ \emph {et~al.}(2018)\citenamefont {Wu},
  \citenamefont {Fatemi}, \citenamefont {Gibson}, \citenamefont {Watanabe},
  \citenamefont {Taniguchi}, \citenamefont {Cava},\ and\ \citenamefont
  {Jarillo-Herrero}}]{Wu2018}%
  \BibitemOpen
  \bibfield  {author} {\bibinfo {author} {\bibfnamefont {S.}~\bibnamefont
  {Wu}}, \bibinfo {author} {\bibfnamefont {V.}~\bibnamefont {Fatemi}}, \bibinfo
  {author} {\bibfnamefont {Q.~D.}\ \bibnamefont {Gibson}}, \bibinfo {author}
  {\bibfnamefont {K.}~\bibnamefont {Watanabe}}, \bibinfo {author}
  {\bibfnamefont {T.}~\bibnamefont {Taniguchi}}, \bibinfo {author}
  {\bibfnamefont {R.~J.}\ \bibnamefont {Cava}}, \ and\ \bibinfo {author}
  {\bibfnamefont {P.}~\bibnamefont {Jarillo-Herrero}},\ }\href@noop {}
  {\bibfield  {journal} {\bibinfo  {journal} {Science}\ }\textbf {\bibinfo
  {volume} {359}},\ \bibinfo {pages} {76} (\bibinfo {year} {2018})}\BibitemShut
  {NoStop}%
\bibitem [{\citenamefont {P{\'e}terfalvi}\ \emph {et~al.}(2015)\citenamefont
  {P{\'e}terfalvi}, \citenamefont {Korm{\'a}nyos},\ and\ \citenamefont
  {Burkard}}]{bib:Korman}%
  \BibitemOpen
  \bibfield  {author} {\bibinfo {author} {\bibfnamefont {C.~G.}\ \bibnamefont
  {P{\'e}terfalvi}}, \bibinfo {author} {\bibfnamefont {A.}~\bibnamefont
  {Korm{\'a}nyos}}, \ and\ \bibinfo {author} {\bibfnamefont {G.}~\bibnamefont
  {Burkard}},\ }\href@noop {} {\bibfield  {journal} {\bibinfo  {journal}
  {Physical Review B}\ }\textbf {\bibinfo {volume} {92}},\ \bibinfo {pages}
  {245443} (\bibinfo {year} {2015})}\BibitemShut {NoStop}%
\bibitem [{\citenamefont {Enaldiev}(2017)}]{Enaldiev}%
  \BibitemOpen
  \bibfield  {author} {\bibinfo {author} {\bibfnamefont {V.~V.}\ \bibnamefont
  {Enaldiev}},\ }\href@noop {} {\bibfield  {journal} {\bibinfo  {journal}
  {Phys. Rev. B}\ }\textbf {\bibinfo {volume} {96}},\ \bibinfo {pages} {235429}
  (\bibinfo {year} {2017})}\BibitemShut {NoStop}%
\bibitem [{\citenamefont {Slobodeniuk}\ and\ \citenamefont
  {Basko}(2016)}]{Basko}%
  \BibitemOpen
  \bibfield  {author} {\bibinfo {author} {\bibfnamefont {A.}~\bibnamefont
  {Slobodeniuk}}\ and\ \bibinfo {author} {\bibfnamefont {D.}~\bibnamefont
  {Basko}},\ }\href@noop {} {\bibfield  {journal} {\bibinfo  {journal} {2D
  Materials}\ }\textbf {\bibinfo {volume} {3}},\ \bibinfo {pages} {035009}
  (\bibinfo {year} {2016})}\BibitemShut {NoStop}%
\bibitem [{\citenamefont {Korm{\'a}nyos}\ \emph {et~al.}(2014)\citenamefont
  {Korm{\'a}nyos}, \citenamefont {Z{\'o}lyomi}, \citenamefont {Drummond},\ and\
  \citenamefont {Burkard}}]{kormanyos2014}%
  \BibitemOpen
  \bibfield  {author} {\bibinfo {author} {\bibfnamefont {A.}~\bibnamefont
  {Korm{\'a}nyos}}, \bibinfo {author} {\bibfnamefont {V.}~\bibnamefont
  {Z{\'o}lyomi}}, \bibinfo {author} {\bibfnamefont {N.~D.}\ \bibnamefont
  {Drummond}}, \ and\ \bibinfo {author} {\bibfnamefont {G.}~\bibnamefont
  {Burkard}},\ }\href@noop {} {\bibfield  {journal} {\bibinfo  {journal}
  {Physical Review X}\ }\textbf {\bibinfo {volume} {4}},\ \bibinfo {pages}
  {011034} (\bibinfo {year} {2014})}\BibitemShut {NoStop}%
\bibitem [{\citenamefont {Ochoa}\ and\ \citenamefont
  {Rold{\'a}n}(2013)}]{Roldan}%
  \BibitemOpen
  \bibfield  {author} {\bibinfo {author} {\bibfnamefont {H.}~\bibnamefont
  {Ochoa}}\ and\ \bibinfo {author} {\bibfnamefont {R.}~\bibnamefont
  {Rold{\'a}n}},\ }\href@noop {} {\bibfield  {journal} {\bibinfo  {journal}
  {Physical Review B}\ }\textbf {\bibinfo {volume} {87}},\ \bibinfo {pages}
  {245421} (\bibinfo {year} {2013})}\BibitemShut {NoStop}%
\bibitem [{\citenamefont {Van Der~Zande}\ \emph {et~al.}(2013)\citenamefont
  {Van Der~Zande}, \citenamefont {Huang}, \citenamefont {Chenet}, \citenamefont
  {Berkelbach}, \citenamefont {You}, \citenamefont {Lee}, \citenamefont
  {Heinz}, \citenamefont {Reichman}, \citenamefont {Muller},\ and\
  \citenamefont {Hone}}]{Zande}%
  \BibitemOpen
  \bibfield  {author} {\bibinfo {author} {\bibfnamefont {A.~M.}\ \bibnamefont
  {Van Der~Zande}}, \bibinfo {author} {\bibfnamefont {P.~Y.}\ \bibnamefont
  {Huang}}, \bibinfo {author} {\bibfnamefont {D.~A.}\ \bibnamefont {Chenet}},
  \bibinfo {author} {\bibfnamefont {T.~C.}\ \bibnamefont {Berkelbach}},
  \bibinfo {author} {\bibfnamefont {Y.}~\bibnamefont {You}}, \bibinfo {author}
  {\bibfnamefont {G.-H.}\ \bibnamefont {Lee}}, \bibinfo {author} {\bibfnamefont
  {T.~F.}\ \bibnamefont {Heinz}}, \bibinfo {author} {\bibfnamefont {D.~R.}\
  \bibnamefont {Reichman}}, \bibinfo {author} {\bibfnamefont {D.~A.}\
  \bibnamefont {Muller}}, \ and\ \bibinfo {author} {\bibfnamefont {J.~C.}\
  \bibnamefont {Hone}},\ }\href@noop {} {\bibfield  {journal} {\bibinfo
  {journal} {Nature materials}\ }\textbf {\bibinfo {volume} {12}},\ \bibinfo
  {pages} {554} (\bibinfo {year} {2013})}\BibitemShut {NoStop}%
\bibitem [{SM()}]{SM}%
  \BibitemOpen
  \href@noop {} {\bibinfo  {journal} {See Supplemental Material at ...
  expressions for edge state functions and spectra via the model parameters,
  and derivation of
  Eqs.(\ref{RPA_gen}),(\ref{intra_dispeq}),(\ref{inter_disp})}\ }\BibitemShut
  {NoStop}%
\bibitem [{\citenamefont {Rostami}\ \emph {et~al.}(2016)\citenamefont
  {Rostami}, \citenamefont {Asgari},\ and\ \citenamefont
  {Guinea}}]{bib:Guinea}%
  \BibitemOpen
\bibfield  {journal} {  }\bibfield  {author} {\bibinfo {author} {\bibfnamefont
  {H.}~\bibnamefont {Rostami}}, \bibinfo {author} {\bibfnamefont
  {R.}~\bibnamefont {Asgari}}, \ and\ \bibinfo {author} {\bibfnamefont
  {F.}~\bibnamefont {Guinea}},\ }\href@noop {} {\bibfield  {journal} {\bibinfo
  {journal} {Journal of Physics: Condensed Matter}\ }\textbf {\bibinfo {volume}
  {28}},\ \bibinfo {pages} {495001} (\bibinfo {year} {2016})}\BibitemShut
  {NoStop}%
\bibitem [{\citenamefont {Volkov}\ and\ \citenamefont
  {Zagorodnev}(2013)}]{Volkov_Zag}%
  \BibitemOpen
  \bibfield  {author} {\bibinfo {author} {\bibfnamefont {V.~A.}\ \bibnamefont
  {Volkov}}\ and\ \bibinfo {author} {\bibfnamefont {I.}~\bibnamefont
  {Zagorodnev}},\ }\href@noop {} {\bibfield  {journal} {\bibinfo  {journal}
  {JETP letters}\ }\textbf {\bibinfo {volume} {97}},\ \bibinfo {pages} {404}
  (\bibinfo {year} {2013})}\BibitemShut {NoStop}%
\bibitem [{\citenamefont {Das~Sarma}\ and\ \citenamefont
  {Hwang}(2009)}]{DasSarma}%
  \BibitemOpen
  \bibfield  {author} {\bibinfo {author} {\bibfnamefont {S.}~\bibnamefont
  {Das~Sarma}}\ and\ \bibinfo {author} {\bibfnamefont {E.~H.}\ \bibnamefont
  {Hwang}},\ }\href@noop {} {\bibfield  {journal} {\bibinfo  {journal} {Phys.
  Rev. Lett.}\ }\textbf {\bibinfo {volume} {102}},\ \bibinfo {pages} {206412}
  (\bibinfo {year} {2009})}\BibitemShut {NoStop}%
\bibitem [{\citenamefont {Wendler}\ and\ \citenamefont
  {Grigoryan}(1994)}]{Wendler1994}%
  \BibitemOpen
  \bibfield  {author} {\bibinfo {author} {\bibfnamefont {L.}~\bibnamefont
  {Wendler}}\ and\ \bibinfo {author} {\bibfnamefont {V.~G.}\ \bibnamefont
  {Grigoryan}},\ }\href@noop {} {\bibfield  {journal} {\bibinfo  {journal}
  {Physical Review B}\ }\textbf {\bibinfo {volume} {49}},\ \bibinfo {pages}
  {14531} (\bibinfo {year} {1994})}\BibitemShut {NoStop}%
\bibitem [{\citenamefont {Andersen}\ \emph {et~al.}(2014)\citenamefont
  {Andersen}, \citenamefont {Jacobsen},\ and\ \citenamefont
  {Thygesen}}]{Andersen}%
  \BibitemOpen
  \bibfield  {author} {\bibinfo {author} {\bibfnamefont {K.}~\bibnamefont
  {Andersen}}, \bibinfo {author} {\bibfnamefont {K.~W.}\ \bibnamefont
  {Jacobsen}}, \ and\ \bibinfo {author} {\bibfnamefont {K.~S.}\ \bibnamefont
  {Thygesen}},\ }\href@noop {} {\bibfield  {journal} {\bibinfo  {journal}
  {Physical Review B}\ }\textbf {\bibinfo {volume} {90}},\ \bibinfo {pages}
  {161410} (\bibinfo {year} {2014})}\BibitemShut {NoStop}%
\bibitem [{\citenamefont {Vitlina}\ \emph {et~al.}(2010)\citenamefont
  {Vitlina}, \citenamefont {Magarill},\ and\ \citenamefont
  {Chaplik}}]{Chaplik}%
  \BibitemOpen
  \bibfield  {author} {\bibinfo {author} {\bibfnamefont {R.~Z.}\ \bibnamefont
  {Vitlina}}, \bibinfo {author} {\bibfnamefont {L.~I.}\ \bibnamefont
  {Magarill}}, \ and\ \bibinfo {author} {\bibfnamefont {A.~V.}\ \bibnamefont
  {Chaplik}},\ }\href@noop {} {\bibfield  {journal} {\bibinfo  {journal} {JETP
  letters}\ }\textbf {\bibinfo {volume} {92}},\ \bibinfo {pages} {692}
  (\bibinfo {year} {2010})}\BibitemShut {NoStop}%
\end{thebibliography}%

\pagebreak
\begin{center}
\textbf{\large Supplemental Materials for ''Collective excitations in two-component one-dimensional massless Dirac plasma''}
\end{center}
\setcounter{equation}{0}
\setcounter{figure}{0}
\setcounter{table}{0}
\setcounter{page}{1}
\makeatletter
\renewcommand{\theequation}{S\arabic{equation}}
\renewcommand{\thefigure}{S\arabic{figure}}
\renewcommand{\bibnumfmt}[1]{[S#1]}

\section{Spectra and wave functions of Edge States}

In $K$ and $K'$ valleys of sTMD monolayers electrons and hole around band edges are described by the following Hamiltonian \cite{bib:Xiao}:
\small
\begin{equation}\label{SM_H_0}
H_{\tau} = v\left(\tau\sigma_xp_x + \sigma_yp_y\right) +\sigma_z\left[m - \tau \hat{s}_z(\Delta_v-\Delta_c)\right]+ \sigma_0 \tau \hat{s}_z(\Delta_v+\Delta_c), 
\end{equation}
\normalsize
where $\tau=\pm 1$ is the valley index, $\sigma_{x,y,z}$ and $\sigma_0$ are Pauli and identity matrixes, respectively, acting in band subspace, $\hat{s}_z$ is $z$ component of the spin-1/2 operator, $v$ is matrix element of momentum between the band extremum wave functions, $2\Delta_{c,v}$ is atomic spin-orbit splitting in conduction and valence bands, respectively, $2m$ is band gap without the splitting. 

Below in this section we calculate spectra and wave functions of edge states (ESs) for two different types of spin-orbit interaction (SOI) mechanisms.  

\subsection{Bulk SOI}\label{section_BSOI}

First, we consider bulk mechanisms of SOI, such as Bychkov-Rashba SOI with substrate, SOI induced by perpendicular electric field, and SOI induced by in-plane magnetic field. In general, these interactions can be described by a perturbation Hamiltonian:
\begin{equation}\label{SM_H_1}
\delta H_{\tau} = 
\left(
\begin{array}{cc}
0 & V_{\tau} \\
V^{+}_{\tau} & 0
\end{array}
\right).
\end{equation}
Here $V_{\tau}$ is square matrix of the second order, which in case of Bychkov-Rashba interface SOI or SOI caused by perpendicular electric field, has the form \cite{Roldan,Basko}:
\begin{equation}
V_{\tau} = 
\left( 
\begin{array}{cc}
0 & \lambda^*\, \delta_{\tau,-1} \\
-\lambda\, \delta_{\tau,1} & 0
\end{array}
\right),
\end{equation} 
where $\lambda$ is parameter that describes value of the interactions ($\lambda\sim 1$ meV). In case of in-plane magnetic field $V_{\tau}=[g\mu_B/2](B_x-iB_y)\sigma_0$ (\mbox{$g\approx 2$} is effective $g$-factor and $\mu_B$ is the Bohr magneton). For bulk SOI we employ spin-decoupled boundary condition (BC) at the edge:
\begin{equation}\label{SM_BC_1} \left[\psi_{c,\tau}^{\uparrow (\downarrow)} + a_{1,2;\tau} \psi_{v,\tau}^{\uparrow (\downarrow)}\right]_{y=0}=0,
\end{equation}
where $a_{1,\tau}$ ($a_{2,\tau}$) is real phenomenological parameter that characterizes edge properties for spin up (spin down) states. Time reversal symmetry impose the following relation: $a_{1,\tau}=a_{2,-\tau}$ \cite{Enaldiev}. In zero approximation on $\delta H$ (\ref{SM_H_1}), Hamiltonian (\ref{SM_H_0}) with BC (\ref{SM_BC_1}) gives rise to two subbands of ESs in each valley \cite{Enaldiev}:
\begin{equation}\label{SM_ES_spectra}
\varepsilon_{n,p}^{(\tau)} = -\tau v_{n}^{(\tau)}p + \Delta_{n}^{(\tau)},
\end{equation}
here \mbox{$n=1,2$} is the ES subband index, \mbox{$v_{n}^{(\tau)} = 2a_{n,\tau}v/(1 + a^2_{n,\tau})$} is ES velocity and 
\begin{equation}
\Delta_{n}^{(\tau)} = m\frac{a^2_{n,\tau}-1}{a^2_{n,\tau} + 1} + (-1)^{n+1}\tau\frac{\Delta_v + a^2_{n,\tau} \Delta_c}{1 + a^2_{n,\tau}} 
\end{equation}
is value of e-h asymmetry. ES subbands possess linear dispersion and are splitted in energy as $\Delta^{(\tau)}_{1}\neq\Delta^{(\tau)}_{2}$. Zero approximation wave functions of the ESs have the following form: 
\begin{eqnarray}\label{SM_ES_WF}
\psi^{(0)}_{1,p,\tau} = C_{1,p,\tau}
\left( 1, -1/a_{1,\tau}, 0, 0 \right)^{\rm T} e^{-\kappa_{1,p,\tau}y + ipx}, \\ \psi^{(0)}_{2,p,\tau} = C_{2,p,\tau}
\left( 0, 0, 1, -1/a_{2,\tau}\right)^{\rm T} e^{-\kappa_{2,p,\tau}y + ipx}, 
\end{eqnarray}
where $C_{n,p,\tau} = \left(2\kappa_{n,p,\tau}a^2_{n,\tau}/(1 + a^2_{n,\tau} )L_x\right)^{1/2}$ is normalization factor, and 
\begin{equation}\label{SM_decay_length}
\hbar\kappa_{n,p,\tau} = -\tau p \dfrac{1 - a_{n,\tau}^2}{1 + a_{n,\tau}^2} +  \frac{ 2a_{n,\tau} }{1 + a^2_{n,\tau} } \left(\frac{m}{v} -(-1)^{n+1}\tau \frac{\Delta_v - \Delta_c}{2v} \right)
\end{equation}
is inverse decay length of ESs. ESs exist for those quasimomenta while their inverse decay length (\ref{SM_decay_length}) is positive. In Ref.\cite{Enaldiev} was shown that ESs (\ref{SM_ES_WF}),(\ref{SM_ES_spectra}) are in the gap at $a_{1,2;1}>0$ ($a_{1,2;-1}<0$). We also assume that $a_{\pm 1,\tau}\sim 1$ for ESs to be in the gap at small enough quasimomenta. Because of quite large band gap, the inverse decay length (\ref{SM_decay_length}) weakly depends on the quasimomentum of the ESs. Therefore, below we take its value at Fermi-level at calculation of plasmon spectra. 

Perturbation (\ref{SM_H_1}) couples ESs belonging to different subbands. Since ES subbands are non degenerate, only their wave functions are changed in the first approximation on $\delta H_{\tau}$ (\ref{SM_H_1}):
\begin{equation}\label{SM_WF_bulk_perturb_1}
\psi_{1,p,\tau} = \psi^{(0)}_{1,p,\tau} + \frac{\langle\psi_{2,p,\tau}^{(0)}|\delta H_{\tau}|\psi_{1,p,\tau}^{(0)}\rangle}{\varepsilon_{1,p}^{(\tau)} - \varepsilon_{2,p}^{(\tau)}}\psi^{(0)}_{2,p,\tau},
\end{equation}

\begin{equation}\label{SM_WF_bulk_perturb_2}
\psi_{2,p,\tau} = \psi^{(0)}_{2,p,\tau} + \frac{\langle\psi_{1,p,\tau}^{(0)}|\delta H_{\tau}|\psi_{2,p,\tau}^{(0)}\rangle}{\varepsilon_{2,p}^{(\tau)} - \varepsilon_{1,p}^{(\tau)}}\psi^{(0)}_{1,p,\tau}.
\end{equation}
 
\subsection{Edge SOI}\label{section_ESOI}

In this section we consider the second type of SOI, that is caused by breaking of crystal potential in perpendicular to edge direction. Such SOI can be described by a spin-coupled BC \cite{Enaldiev}:
\begin{equation}\label{SM_BC_2}
\left[\Psi^{\uparrow}_{\tau} - M_{\tau}\Psi^{\downarrow}_{\tau} \right]_{y=0} = 0,
\end{equation}
where $\Psi^{\uparrow(\downarrow)}_{\tau} = (\psi_{c,\tau}^{\uparrow (\downarrow)},\psi_{v,\tau}^{\uparrow (\downarrow)})$ is two-component envelope functions, and $M$ is the second order matrix with elements:
\begin{align}
m_{11} = i(\sinh\xi + \tau\cosh\xi\cosh\eta\cos\nu)\\
m_{22} = i(\sinh\xi - \tau\cosh\xi\cosh\eta\cos\nu)\\
m_{12} = i\tau\cosh\xi(\sinh\eta - \cosh\eta\sin\nu) \\
m_{21} = -i\tau\cosh\xi(\sinh\eta + \cosh\eta\sin\nu),
\end{align}
where $\xi,\eta,\nu$ are phenomenological parameters that describe SOI at the edge. In case of the edge SOI, there also exists two subbands of ESs which are described by the following wave functions:
\begin{equation}\label{SM_WF_ESOI}
\psi_{n,p,\tau} = C_{n,p,\tau}
\left[
\begin{array}{c}
M_{\tau}
\left(
\begin{array}{c}
1 \\
G^{\downarrow}_{n,p,\tau}
\end{array}
\right)e^{\kappa^{\uparrow}_{n,p,\tau}y} \\
\left(
\begin{array}{c}
1 \\
G^{\downarrow}_{n,p,\tau}
\end{array}
\right)e^{\kappa^{\downarrow}_{n,p,\tau}y}
\end{array}
\right]e^{ipx},
\end{equation}
where 
\begin{equation}
G^{\uparrow(\downarrow)}_{n,p,\tau} = \frac{v(\tau p - \hbar\kappa^{\uparrow(\downarrow)}_{n,p,\tau})}{\varepsilon_{n,p}^{(\tau)}+m\mp\tau\Delta_v}\nonumber
\end{equation}
and inverse decay lengthes are expressed as follows:
\begin{equation}\label{SM_decay_length_ESOI}
\hbar\kappa^{\uparrow(\downarrow)}_{n,p,\tau} = \sqrt{p^2-(\varepsilon_{n,p}^{(\tau)} - m \mp\tau\Delta_c)(\varepsilon_{n,p}^{(\tau)} + m \mp\tau\Delta_v)/v^2}.
\end{equation}
Energies of ES subbands are determined by a dispersion equation:
\begin{align}\label{SM_disp_eq_ESOI}
\left(1 + a_{1,\tau}G^{\uparrow}_{n,p,\tau}\right)\left(1 + a_{2,\tau}G^{\downarrow}_{n,p,\tau}\right) + \nonumber\\ 
\frac{\tau v(\tanh\xi - 1)}{\sinh\eta + \cosh\eta\sin\nu}\left(G^{\uparrow}_{n,p,\tau} - G^{\downarrow}_{n,p,\tau}\right)=0, 
\end{align}
where 
\begin{equation}\label{SM_a_12}
a_{1,2;\tau}=\pm\tau \frac{1 \pm\tau \cosh\eta\cos\nu}{\sinh\eta+\cosh\eta\sin\nu}
\end{equation}
are expressions for the introduced previously section boundary parameters via $\eta,\nu$. Coupling of the ES subbands by the edge SOI is provided by the last term in Eq.(\ref{SM_disp_eq_ESOI}). Taking this term as a perturbation, we obtain the same spectra of ESs as in case of the bulk SOI (\ref{SM_ES_spectra}), but with renormalized e-h asymmetry term:
\begin{align}
\widetilde{\Delta}_{n}^{(\tau)} = \Delta_{n}^{(\tau)} + 
\frac{(-1)^{n+1}2\tau\Delta_v(\tanh\xi - 1)}{(1+a_{n,\tau}^2)\left(1 + \frac{\Delta_v (\tau(-1)^{n+1}+\cosh\eta\sin\nu)}{\varepsilon_{n,0}^{(\tau)} + m - (-1)^{n+1}\tau\Delta_v}\right)} 
\end{align}
As it follows from Eq.(\ref{SM_a_12}) \mbox{$a_{1,1}>a_{2,1}$}. This means that ESs from the upper subband always has greater absolute value of velocity than from the lower one. Therefore, ES subbands do not intersect each other at any quasimomentum as without SOI as well taking the latter into account. 

\section{Derivation of plasmon dispersion equation in self consistent Random phase approximation}

In this section we derive general dispersion equation for 1D plasmons in self-consistent random phase approximation (RPA), taking into account only transitions between ESs. First, we solve quantum kinetic equation for density matrix $\rho$ in the first order on electric potential:
\begin{equation}\label{SM_QKE}
i\hbar \frac{\partial \rho}{\partial t} = \left[H_0  - e\left(\varphi_q(y)e^{iqx-i(\omega+i0)t}+ c.c.\right), \rho\right]
\end{equation}
where $H_0=H_{\tau} + \delta H_{\tau}$ with BC (\ref{SM_BC_1}) for ES wave function set (\ref{SM_WF_bulk_perturb_1}), (\ref{SM_WF_bulk_perturb_2}), or $H_0=H_{\tau}$ with BC (\ref{SM_BC_2}) for ES wave function set (\ref{SM_WF_ESOI}). The first order amendment to density matrix reads as follows:
\begin{equation}\label{SM_DM_1}
\rho^{(1)}_{\lambda,\lambda'} = -e\langle\lambda|\varphi_q(y)e^{iqx}|\lambda'\rangle\delta_{\tau_{\lambda},\tau_{\lambda'}} \frac{f(\varepsilon_{\lambda'}) - f(\varepsilon_{\lambda})}{\varepsilon_{\lambda'} - \varepsilon_{\lambda} + \hbar\omega+ i0},
\end{equation}
where $e$ is absolute value of electron charge, $\lambda$ is set of ES quantum numbers, $f(\varepsilon)$ is Fermi-Dirac distribution function. Fourier harmonics of induced electron density is expressed in the following way: 
\begin{equation}\label{SM_density}
n^{(1)}_{q}(y_e) = \frac{-e}{L_x}\sum_{\lambda,\lambda'}\rho^{(1)}_{\lambda,\lambda'}\langle \lambda'|\delta(y-y_e)e^{-iqx}|\lambda\rangle
\end{equation}

Inserting (\ref{SM_density}) in the Poisson equation we obtain a closed equation for the electric potential harmonics:
\begin{equation}\label{SM_Poisson}
\varphi_q(y)e^{iqx} = \frac{2}{\epsilon^*}\int_{0}^{+\infty}K_0\left(\left|q(y-y')\right|\right)n_q^{(1)}(y')e^{iqx}dy', 
\end{equation}  
where $K_0(x)$ is the Macdonald function describing Coulomb interaction in one dimension, $\epsilon^*$ is dielectric constant of environment. Substituting (\ref{SM_density}) in the previous equation and taking matrix element $\langle \mu'|\dots|\mu\rangle \delta_{\tau_{\mu}\tau_{\mu'}}$ from both side of the equation, we arrive to the final expression used in the main text:
\begin{eqnarray}\label{SM_RPA_gen}
\langle \mu'|\varphi_{q}(y)e^{iqx}|\mu\rangle\delta_{\tau_{\mu'}\tau_{\mu}} = \qquad\qquad\qquad\qquad\qquad\qquad\qquad\nonumber\\
 \frac{2e^2}{\epsilon^*L_x}\sum_{\lambda\lambda'}\frac{f(\varepsilon_{\lambda'}) - f(\varepsilon_{\lambda}) }{ \varepsilon_{\lambda'} - \varepsilon_{\lambda} + \hbar\omega +i0 }\langle \lambda|\varphi_{q}(y)e^{iqx}|\lambda'\rangle\delta_{\tau_{\lambda'}\tau_{\lambda}}\times \nonumber\\ 
\langle\lambda'|\langle\mu'|e^{iq(x-x')}K_0\left(|q(y-y')|\right)|\mu\rangle|\lambda\rangle \qquad\quad
\end{eqnarray}
 
\subsection{Dispersion equation for intrasubband plasmons}

Leaving only diagonal matrix elements on the subband number in Eq.(\ref{SM_RPA_gen}) (f.e. \mbox{$|\mu\rangle=|1,p',1\rangle $} and \mbox{$|\mu'\rangle =|1, p'+q,1\rangle$}) we arrive to the following equation:
\small
\begin{widetext}
\begin{equation}\label{SM_RPA_intra}
\begin{split}
\langle 1,p'+q,1\left|\varphi_{q}(y)e^{iqx}\right|1,p',1\rangle = \qquad\qquad\qquad\qquad\qquad\qquad\qquad\qquad\qquad\qquad\qquad\qquad\qquad\qquad\qquad\qquad\qquad\qquad\qquad\qquad 
\\ 
\frac{2e^2}{\epsilon^*L_x}\sum_{p,\tau}
\frac{ f(\varepsilon_{1,p}^{(\tau)}) - f(\varepsilon_{1,p+q}^{(\tau)}) }{ \varepsilon_{1,p}^{(\tau)} - \varepsilon_{1,p+q}^{(\tau)} +\hbar\omega + i0} \langle 1,p+q,\tau\left|\varphi_{q}(y)e^{iqx}\right|1,p,\tau\rangle 
\langle 1,p'+q,1|\langle 1,p,\tau\left|K_0(\left|q(y-y')\right|)e^{iq(x-x')}\right|1,p+q,\tau\rangle|1,p',1\rangle +  \\
\frac{2e^2}{\epsilon^*L_x}\sum_{p,\tau}
\frac{ f(\varepsilon_{2,p}^{(\tau)}) - f(\varepsilon_{2,p+q}^{(\tau)}) }{ \varepsilon_{2,p}^{(\tau)} - \varepsilon_{2,p+q}^{(\tau)} +\hbar\omega + i0 } \langle 2,p+q,\tau\left|\varphi_{q}(y)e^{iqx}\right|2,p,\tau\rangle 
\langle 1,p'+q, 1|\langle 2,p,\tau\left|K_0(\left|q(y-y')\right|)e^{iq(x-x')}\right|2,p+q,\tau\rangle|1,p',1\rangle.  
\end{split}
\end{equation}
\end{widetext}
\normalsize
As ESs are exponentially localized in vicinity of the edge, the diagonal matrix elements of electric potential in the long wavelength limit $q\kappa^{-1}\ll 1$ ($\kappa$ is inverse decay length of ESs at Fermi-level) can be estimated as:
\begin{eqnarray}\label{SM_potential_intra}
\langle 1,p+q,\tau|\varphi_{q}(y)e^{iqx}|1,p,\tau\rangle = \nonumber \qquad\qquad\qquad\\
\langle 2,p+q,\tau|\varphi_{q}(y)e^{iqx}|2,p,\tau\rangle \approx \varphi_{q}(0).
\end{eqnarray}
Taking asymptotic expression for 1D Coulomb potential $K_0(x)\approx \ln(2/x\zeta)$ in the long wavelength limit ($\zeta=e^{\gamma}$, $\gamma=0.577\dots$ is the Euler constant), we evaluate its matrix elements as follows:
\small
\begin{equation}\label{SM_Coulomb_intra_1}
\begin{split}
\langle 1,p'+q,1|\langle 1,p,\tau\left|K_0(\left|q(y-y')\right|)e^{iq(x-x')}\right|1,p+q,\tau\rangle|1,p',\tau\rangle \approx \qquad\\
 \ln\left[\frac{2\left(\kappa_{1,p,\tau} +\kappa_{1,p',1}\right) }{q\zeta}\right]\approx \ln\left[\frac{4\kappa_{1,p_{F_1},1}}{q\zeta}\right], \qquad\qquad\qquad\qquad\qquad
\end{split}  
\end{equation}
\begin{equation}\label{SM_Coulomb_intra_2}
\begin{split}
\langle 1,p'+q,1|\langle 2,p,\tau\left|K_0(\left|q(y-y')\right|)e^{iq(x-x')}\right|2,{\small p+q},\tau\rangle|1,p',1\rangle \approx \qquad\qquad\qquad\\
\ln\left[\frac{2\left(\kappa_{2,p,\tau} +\kappa_{1,p', 1}\right) }{q\zeta}\right]\approx \ln\left[\frac{2\left( \kappa_{2,p_{F_2},1} + \kappa_{1,p_{F_1},1}\right)}{q\zeta}\right] \approx \qquad\qquad\qquad\qquad\quad \\ \ln\left[\frac{4\kappa_{2,p_{F_2},1}}{q\zeta}\right], \qquad\qquad\qquad\qquad\quad
\end{split}
\end{equation}  
\normalsize
Since the inverse decay lengths (\ref{SM_decay_length}) or (\ref{SM_decay_length_ESOI}) is of weak dependence on quasimomentum, we take its value at corresponding Fermi-level. Further, due to smoothness of logarithm function, we will put $\ln(4\kappa_{1,p_{F_1},1}/q\zeta)\approx\ln(4\kappa_{2,p_{F_2},1}/q\zeta)$ in the matrix elements (\ref{SM_Coulomb_intra_1}),(\ref{SM_Coulomb_intra_2}).  Inserting Eqs.(\ref{SM_potential_intra})-(\ref{SM_Coulomb_intra_2}) in the system (\ref{SM_RPA_intra}) we finally obtain the dispersion equation for intrasubband plasmons that is used in the main text:
\begin{equation}
\begin{split}
1-\frac{2e^2}{\epsilon^*}\ln\left(\frac{4\kappa_{2,p_{F_2},1}}{|q|\zeta}\right)\sum_{\tau}\left[\Pi^{(\tau)}_{11}(\omega, q) + \Pi^{(\tau)}_{22}(\omega, q)\right]=0,\\
\Pi^{(\tau)}_{n'n}(\omega, q)=\frac{1}{L_x}\sum_{p}\frac{f\left(\varepsilon_{n',p}^{(\tau)}\right) - f\left(\varepsilon_{n,p+q}^{(\tau)}\right) }{ \varepsilon_{n',p}^{(\tau)} - \varepsilon_{n,p+q}^{(\tau)} + \hbar\omega +i0 }, \, n,n'=\{1,2\} \\
\\
\\
\\
\end{split}
\end{equation} 
Therefore, in the main text $\kappa_{\rm intra}=\kappa_{2,p_{F_2},1}\approx \kappa_{1, p_{F_1},1}$.

\subsection{Dispersion equation for intersubband plasmons}

\begin{figure}
\includegraphics{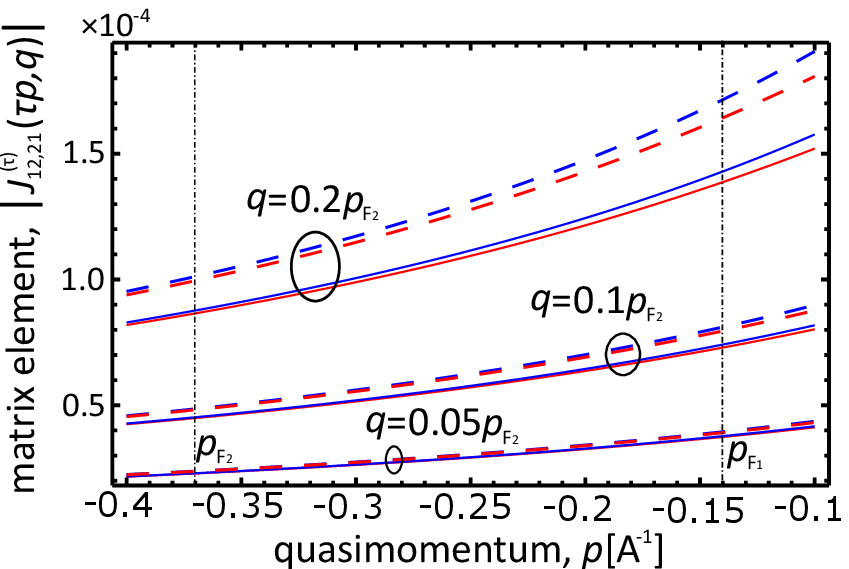}
\caption{Quasimomentum dependence of intersuband matrix element absolute values $|J_{12}^{(\tau)}(\tau p,q)|$ (blue), $|J_{21}^{(\tau)}(\tau p,q)|$ (red) in case of bulk SOI at three values of wave vectors $q=0.05p_{F_2}0.1p_{F_2},0.2p_{F_2}$. Solid (dashed) lines respond to the $K$ ($K'$) valley. Vertical line shows quasimomenta corresponding to Fermi-level in each edge subband (see, Fig.\ref{Fig1} of the main text). Calculation was carried out at: $2m=1.8$ eV, $v=2.5$ eV$\cdot$A, $2\Delta_c=3$ meV, $2\Delta_v=0.148$ eV, $a_{1,1}=a_{2,-1}=0.5$, $a_{2,1}=a_{1,-1}=0.35$, $|\lambda|=1$meV.
\label{SM_J_BSOI}}
\end{figure}

\begin{figure}
\includegraphics{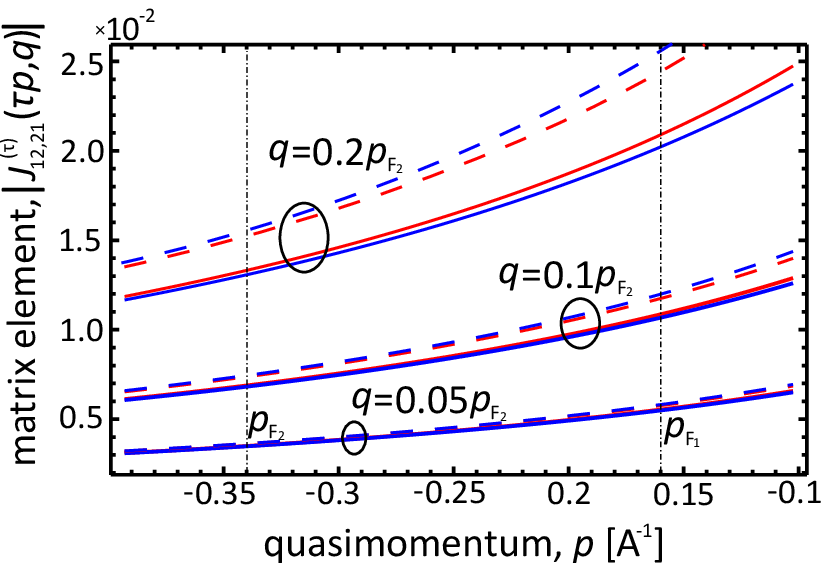}
\caption{Quasimomentum dependence of intersuband matrix element absolute values $|J_{12}^{(\tau)}(\tau p,q)|$ (blue), $|J_{21}^{(\tau)}(\tau p,q)|$ (red) in case of edge SOI at three values of wave vectors $q=0.05p_{F_2}0.1p_{F_2},0.2p_{F_2}$. Solid (dashed) lines respond to the $K$ ($K'$) valley. Vertical line shows quasimomenta corresponding to Fermi-level in each edge subband (see, Fig.\ref{Fig1} of the main text). Calculation was carried out at: $2m=1.8$ eV, $v=2.5$ eV$\cdot$A, $2\Delta_c=3$ meV, $2\Delta_v=0.148$ eV, $\xi=0.2$, $\eta=2.76$, $\nu=0.77$. \label{SM_J_ESOI}}
\end{figure}

Taking \mbox{$|\mu'\rangle=|1,p'+q,1\rangle$}, \mbox{$|\mu\rangle=|2,p',1\rangle$} and keeping only intersubband matrix elements in Eq.(\ref{SM_RPA_gen}) we obtain the following equation:
\small
\begin{widetext}
\begin{equation}\label{SM_RPA_inter}
\begin{split}
\langle 1,p'+q, 1\left|\varphi_{q}(y)e^{iqx}\right|2, p', 1\rangle = \qquad\qquad\qquad\qquad\qquad\qquad\qquad\qquad\qquad\qquad\qquad\qquad\qquad\qquad\qquad\qquad\qquad\qquad\qquad\qquad\qquad  \\
\frac{2e^2}{\epsilon^*L_x}\sum_{p,\tau}
\frac{ f(\varepsilon_{1,p}^{(\tau)}) - f(\varepsilon_{2,p+q}^{(\tau)}) }{ \varepsilon_{1,p}^{(\tau)} - \varepsilon_{2,p+q}^{(\tau)} +\hbar\omega + i0 } \langle 2,p+q,\tau\left|\varphi_{q}(y)e^{iqx}\right|1,p,\tau\rangle 
\langle 1,p'+q, 1|\langle 1,p,\tau\left|K_0(\left|q(y-y')\right|)e^{iq(x-x')}\right|2,p+q,\tau\rangle|2,p', 1\rangle +  \\
\frac{2e^2}{\epsilon^*L_x}\sum_{p,\tau}
\frac{ f(\varepsilon_{2,p}^{(\tau)}) - f(\varepsilon_{1,p+q}^{(\tau)}) }{ \varepsilon_{2,p}^{(\tau)} - \varepsilon_{1,p+q}^{(\tau)} +\hbar\omega + i0 } \langle 1,p+q,\tau\left|\varphi_{q}(y)e^{iqx}\right|2,p,\tau\rangle
\langle 1,p'+q,1|\langle 2,p,\tau\left|K_0(\left|q(y-y')\right|)e^{iq(x-x')}\right|1,p+q,\tau\rangle|2,p',1\rangle.  
\end{split}
\end{equation}
\end{widetext}
\normalsize
In long wavelength limit intersubband matrix elements of electric potential are evaluated as follows:
\small
\begin{equation}\label{SM_potential_inter_1}
\begin{split}
\langle 1,p+q,\tau\left|\varphi_{q}(y)e^{iqx}\right|2,p,\tau\rangle \approx \varphi_{q}(0)\langle 1,p+q,\tau\left|e^{iqx}\right|2,p,\tau\rangle \equiv \\
 \varphi_{q}(0)J_{12}^{(\tau)}(p,q),
\end{split}
\end{equation}
\begin{equation}\label{SM_potential_inter_2}
\begin{split}
\langle 2,p+q,\tau\left|\varphi_{q}(y)e^{iqx}\right|1,p,\tau\rangle\approx \varphi_{q}(0)\langle 2,p+q,\tau\left|e^{iqx}\right|1,p,\tau\rangle \equiv \\
 \varphi_{q}(0)J_{21}^{(\tau)}(p,q),
\end{split}
\end{equation}
\normalsize
where the last equality is in (\ref{SM_potential_inter_1}), (\ref{SM_potential_inter_2}) is a definition of $J_{12,21}^{(\tau)}(p,q)$. In the long wavelength limit intersubband matrix elements of 1D Coulomb potential read as:
\small
\begin{equation}\label{SM_Coulomb_inter}
\begin{split}
\langle 1,p'+q, 1|\langle 1,p,\tau\left|K_0(\left|q(y-y')\right|)e^{iq(x-x')}\right|2,p+q,\tau\rangle|2,p', 1\rangle \approx\\
\ln\left[\frac{2(\kappa_{1,p_{F_2},1}+\kappa_{2,p_{F_2},1})}{q\zeta}\right]J_{12}^{(1)}(p',q)J_{21}^{(\tau)*}(p,q) \qquad\qquad\qquad \\
\langle 1,p'+q, 1|\langle 2, p, \tau \left|K_0(\left|q(y-y')\right|)e^{iq(x-x')}\right|1,p+q,\tau\rangle|2, p', 1\rangle \approx\\
\ln\left[\frac{2(\kappa_{1,p_{F_2}, 1}+\kappa_{2,p_{F_2},1})}{q\zeta}\right]J_{12}^{(1)}(p',q)J_{12}^{(\tau)*}(p,q) \qquad\qquad\qquad \\
\end{split}
\end{equation} 
\normalsize
where we took the inverse decay lengths at $p=p_{F_2}$. In the long wavelength limit matrix elements $J_{12,21}^{(\tau)}(p,q)$ are proportional to wave vector $q$, as at \mbox{$q=0$} ES wave functions belonging different subbands are orthogonal to each other. In the case of edge SOI the matrix elements satisfy symmetry condition: \mbox{$J_{12}^{(1)}(p,q,1)=J_{21}^{(-1)}(-p,-q,-1)$}. But, for the bulk SOI (\ref{SM_H_1}) the condition is following one: \mbox{$J_{12}^{(1)}(p,q,1)=-J_{21}^{(-1)}(-p,-q,-1)$}. Since at small wave vectors the absolute values of the matrix elements approximately equal to each other (i.e. $|J_{12}^{1}(p,q)\approx|J_{21}^{1}(p,q)|\approx|J_{12}^{-1}(-p,q)|\approx|J_{21}^{(-1)}(p,q)|$) and additionally possess weak dependence on quasimomentum (see Figs.\ref{SM_J_BSOI},\ref{SM_J_ESOI}) we treat them as constants taken at $p=p_{F_2}$ in the Eq.(\ref{SM_RPA_inter}). Finally , we obtain the following dispersion equation for intersubband plasmons:
\begin{equation}\label{SM_inter_disp}
\begin{split}
1-\frac{2e^2}{\epsilon^*L_x}\left|J_{12}(p_{F_2},q)\right|^2\ln\left ( \frac{2\left(\kappa_{1,p_{F_2}, 1}+\kappa_{2,p_{F_2},1}\right)}{|q|\zeta} \right )\times \\
\sum_{\tau}\left[\Pi^{(\tau)}_{12}(\omega,q) + \Pi^{(\tau)}_{21}(\omega,q)\right] =0,
\end{split}
\end{equation}
Therefore, in the main text $\kappa_{\rm inter}=(\kappa_{2,p_{F_2},1} + \kappa_{1,p_{F_2},1})/2$.

\end{document}